\documentclass[prd,floatfix,preprintnumbers]{revtex4}
\usepackage{amsmath}
\usepackage{graphicx,epsfig}
\usepackage{color}
\begin{document}
\title{Mapping the Chevallier-Polarski-Linder parametrization
onto Physical Dark Energy Models} 
\author {Robert J. Scherrer}
\affiliation{Department of Physics and Astronomy, Vanderbilt University,
Nashville, TN  ~~37235}

\begin{abstract}
We examine the Chevallier-Polarski-Linder (CPL) parametrization, in
the context of quintessence and barotropic dark energy models, to
determine the subset of such models to which it can provide a good fit.  The CPL parametrization gives
the equation of state parameter $w$ for the dark energy as a linear function of the scale factor $a$, namely
$w = w_0 + w_a(1-a)$.
In the case of quintessence models, we find that over most of the $w_0$, $w_a$ parameter space
the CPL parametrization maps onto a fairly narrow form
of behavior for the potential $V(\phi)$, while a one-dimensional subset of parameter space, for
which $w_a = \kappa (1+w_0)$, with $\kappa$ constant, corresponds
to a wide range of functional forms for $V(\phi)$.  For barotropic models, we show that the functional
dependence of the pressure
on the density, up to a multiplicative constant, depends only on $w_i = w_a + w_0$ and not on $w_0$ and $w_a$ separately.
Our results suggest that the CPL parametrization may not be optimal for testing either type of model.
\end{abstract}

\maketitle

\section{Introduction}

Observations of the accelerated expansion of the universe
\cite{union08,hicken,Amanullah,Union2,Hinshaw,Ade,Betoule}
indicate that approximately
70\% of the energy density in the
universe is in the form of a negative-pressure component,
called dark energy, with the remaining 30\% in the form of nonrelativistic matter (including both baryons
and dark matter).
The dark energy component can be characterized by its equation of state parameter, $w$,
defined as the ratio of the dark energy pressure to its density:
\begin{equation}
\label{w}
w=p/\rho,
\end{equation}
where a cosmological constant, $\Lambda$, corresponds to $w = -1$ and $\rho = constant$.
While a model with a cosmological constant and cold dark matter ($\Lambda$CDM) is consistent
with current observations,
there are many models for dark energy that predict a dynamical equation
of state.
These include, for example, quintessence models, in which the dark energy
arises from a time-dependent scalar field, $\phi$
\cite{Wetterich,RatraPeebles,CaldwellDaveSteinhardt,LiddleScherrer,SteinhardtWangZlatev}.
(See Ref. \cite{Copeland} for a review), and barotropic models, in which the pressure
is simply a prescribed function of the density
\cite{Kamenshchik,Bilic,Bento,linear1,linear2,AB,Quercellini,VDW1,VDW2,LinderScherrer,Bielefeld}.

In practice, an enormous number of dynamical models have been proposed for the dark energy.
Hence, it has been considered useful to classify such models in terms of simple parametrizations
for $w$.  The simplest possibility is to take $w$ to be a constant in time.  However, there are
a variety of problems with the assumption of constant $w$.  With the exception of $w = -1$, a
constant value for $w$ does
not arise naturally in the context of most physically-motivated models.  Further, the assumption
of constant $w$ can provide wildly inaccurate results in the case that $w$ does, in fact, evolve with
time \cite{Maor}.

The next level of complexity is a two-parameter model for $w$ as a function of the scale factor, $a$, or
equivalently, of the redshift, $z$.  By far the most widely-used of such parametrizations is the
Chevallier-Linder-Polarski (CPL) parametrization, which takes $w$ to be a linear function of the scale
factor, namely \cite{CP,Linder}:
\begin{equation}
\label{CPL}
w = w_0 +(1-a)w_a,
\end{equation}
where $w_0$ and $w_a$ are constants.
Other two-parameter approximations have been discussed in the literature \cite{Cooray,Gerke}; for a review
of these and related approaches, see Ref. \cite{SS}.
The CPL parametrization describes fairly gradual evolution from a value of $w=w_0 + w_a$ at early
times to a present-day
value of $w=w_0$.  It has several advantages, not least the fact that it is well-behaved all
of the way from $a=0$ to $a=1$ and, for a variety of models, it can reproduce the predicted
observable quantities (distance or Hubble parameter as a function of redshift) to extraordinarily
high accuracy \cite{Linderlinear1}.  Despite
the variety of proposed parametrizations, the CPL approximation has become the de facto standard
two-parameter description for the evolution of the dark energy equation of state parameter 
(e.g.,
\cite{Union2,Hinshaw,Ade,Betoule}).

Because of its importance, it is worthwhile to examine the applicability of the CPL parametrization
in more detail.
While it is possible to simply treat the CPL parametrization as a convenient heuristic description of the
(unknown) dark energy component, it is also useful to understand the mapping between physical models for dark
energy and the CPL parametrization.  It has been claimed \cite{Linder} that Eq. (\ref{CPL}) provides a good fit
to $w(a)$ for
a wide variety of quintessence models.  In Ref. \cite{ScherrerSen}, it was shown that thawing quintessence
models with a nearly flat potential all converge toward the behavior given
by Eq. (\ref{CPL}), with $w_a \approx -1.5(1+w_0)$.  A similar result
using different techniques was derived in Refs. \cite{Linderlinear1,Linderlinear2},
with $w_a$ given by $w_a \approx -1.58 (1+w_0)$.  This relation is consistent with results
obtained using Monte Carlo sampling
of a variety of different quintessence potentials \cite{Marsh}. A linear relation
between $w_a$ and $1+w_0$ is also evident in the much earlier
work of Kallosh et al. \cite{Kallosh} for the special case of the
linear potential.  On the other hand,
the evolution of $w(a)$ for evolution near a local maximum of the potential (hilltop
quintessence) can deviate strongly
from a linear form \cite{ds1}.

Hence, it is interesting to determine precisely what physical models are well-described by the CPL parametrization, and
which are not.  This is, however, a difficult proposition.  There are an infinite number of
possible quintessence potentials, and even if one could sample the entire space of possibilities, the question
remains of defining a ``good" fit to Eq. (\ref{CPL}).  We therefore take the opposite approach here:  we begin
with Eq. (\ref{CPL}) and determine, for two types of models (quintessence and barotropic fluids) precisely which
models this parametrization maps onto.  This approach has the advantage that one can, in a systematic way, scan over
all possible values of $w_0$ and $w_a$, and then look for patterns in the corresponding physical models.  This mapping
from the CPL parametrization to quintessence models
has been previously explored by Padmanabhan and Choudhury \cite{Pad} and by Guo, Ohta, and Zhang \cite{GOZ},
both of which examined a single pair of $w_0, w_a$ values. Barboza et al. \cite{Barboza} investigated a more
extensive parameter range, but for a different parametrization.
Here we extend this earlier work by scanning over the
full range of $w_0$ and $w_a$ values.  We then perform a similar mapping for barotropic models, which have been
comparatively less well explored.

In the next section we examine the mapping of the CPL parametrization onto quintessence models and determine
the functional form for $V(\phi)$ for this models.  In Sec. III, we perform a similar analysis for barotropic models
and find the pressure as a function of density for these models.  Our conclusions are summarized in Sec. IV.

\section{Quintessence}

In this section, we will consider models in which
the dark energy is provided by a minimally-coupled
scalar field, $\phi$, with equation of motion given by
\begin{equation}
\label{motionq}
\ddot{\phi}+ 3H\dot{\phi} + \frac{dV}{d\phi} =0,
\end{equation}
where the Hubble parameter $H$ is given by
\begin{equation}
\label{H}
H = \left(\frac{\dot{a}}{a}\right) = \sqrt{\rho_T/3},
\end{equation}
and the dot denotes the derivative with respect to time.
Here $a$ is the scale factor (taken to be $a=1$ at the present), $\rho_T$ is the total density, and we work in units
for which $8 \pi G = 1$.
Since we are interested in the evolution of dark energy at relatively
late times, we will consider only the contributions of nonrelativistic
matter (baryons plus dark matter) to $\rho_T$, and ignore, e.g., radiation.

The pressure and density of the
scalar field are given by
\begin{equation}
\label{pphi}
p_\phi = \frac{\dot \phi^2}{2} - V(\phi),
\end{equation}
and
\begin{equation}
\label{rhophi}
\rho_\phi = \frac{\dot \phi^2}{2} + V(\phi),
\end{equation}
respectively, and the equation of state parameter, $w$,
is given by equation (\ref{w}).

Our job is to begin with the CPL parametrization (Eq. \ref{CPL}) and work
backwards to derive $V(\phi)$.  Such a derivation is provided in Refs. \cite{Pad,GOZ,Barboza} (see
also the earlier related discussion in Refs. \cite{Starobinsky,HT}),
so we will simply reproduce their results here.  First, we note that Eqs. (\ref{w}), (\ref{pphi}),
and (\ref{rhophi}) can be combined to yield
\begin{equation}
\label{V(w)}
V = \frac{1}{2}(1-w)\rho_\phi.
\end{equation}
Further, the density of a perfect fluid with equation of state parameter $w$ evolves as
\begin{equation}
\label{rhoevol}
\frac{a}{\rho}\frac{d\rho}{da} = -3(1+w).
\end{equation}
Taking $w(a)$ to be described by the CPL parametrization (Eq. \ref{CPL}), the density evolves as
\begin{equation}
\label{rhoCPL}
\rho = \rho_{\phi 0} a^{-3(1+w_0+w_a)} e^{3w_a(a-1)},
\end{equation}
where $\rho_{\phi 0}$ is the present-day density.  (The $0$ subscript will refer to present-day quantities throughout).
Note that Eqs. (\ref{rhoevol}) and (\ref{rhoCPL}) are not specific to scalar field models; we will use
them again when we discuss barotropic fluids in the next section.

Combining Eqs. (\ref{V(w)}) and (\ref{rhoCPL}) gives an expression for the scalar field potential as a function
of the scale factor:
\begin{equation}
\label{V(a)}
V(a) = \frac{1}{2}\rho_{\phi 0}(1 - w_0 - w_a + w_a a)a^{-3(1+w_0+w_a)}e^{3w_a(a-1)}.
\end{equation}
In order to express $V$ as a function of $\phi$, we need an expression for $\phi(a)$.
From Eqs. (\ref{w}), (\ref{pphi}), and (\ref{rhophi}), we have

\begin{equation}
\frac{d\phi}{dt} = \sqrt{(1+w) \rho_\phi},
\end{equation}
which can be combined with the definition of $H$ (Eq. \ref{H}) to give
\begin{equation}
\label{dphida}
\frac{d\phi}{da} = \frac{\sqrt{(1+w) \rho_\phi}}{aH}.
\end{equation}
Substituting the CPL expression for $w$ into Eq. (\ref{dphida}) and
using $H = \sqrt{(\rho_M + \rho_\phi)/3}$, with $\rho_M = \rho_{M0}a^{-3}$, we get
\begin{equation}
\label{phi(a)}
\phi = \int \frac{\sqrt{3(1+w_0+w_a-w_a a)}}{\sqrt{1+(\rho_{M0}/\rho_{\phi0})
a^{3(w_0+w_a)}e^{3w_a(1-a)}}}\frac{da}{a}.
\end{equation}
Together, Eqs. (\ref{V(a)}) and (\ref{phi(a)}) give a parametric expression for $V(\phi)$, which is our desired result.
Note that Eq. (\ref{phi(a)}) yields an arbitrary additive constant in the expression for $\phi(a)$.  However,
this simply amounts to a shift in the value of $\phi$ by an arbitrary constant
in the expression for $V(\phi)$, which has no physical significance.  We will take $\rho_{M0}/\rho_{\phi 0}
\approx 3/7$ in what follows.

Before exploring the functional form for $V(\phi)$ for given values of $w_0$ and $w_a$, we first consider the
allowed ranges for these two quantities.  Limits derived from observational data have tended
to converge on a fairly narrow ellipse in the $w_0$, $w_a$ plane, with negative slope (i.e., smaller
$w_0$ corresponds to larger $w_a$) \cite{Ade,Betoule}.    However, the validity of these limits is itself dependent
on the assumption that the CPL parametrization is a good fit to the evolution of $w$.  Since
we are interested in probing the extent to which the CPL parametrization describes specific dark energy
models, we will allow $w_0$ and $w_a$ to vary outside of these observational limits.

A more fundamental limit comes from Eqs. (\ref{w}), (\ref{pphi}), and (\ref{rhophi}), which together with
the assumption that $\dot \phi^2/2$ and $V(\phi)$ are both nonnegative, give the well-known
bounds
\begin{equation}
\label{wbound}
-1 \le w \le 1.
\end{equation}
A less stringent limit comes from the requirement that
the evolution correspond to a scalar field rolling downhill in the potential.  While upward-rolling
fields are physically possible, this behavior is normally a transient phenomenon (unless one is
dealing with oscillatory behavior near the minimum of the potential \cite{ds2}).  Hence,
a scalar field rolling uphill in a potential today would correspond to yet another coincidence problem to
add to the well-known approximate equality of matter and dark energy densities at the present.  The requirement
that the field be rolling downhall gives the constraint \cite{Scherrerw}
\begin{equation}
a \frac{dw}{da} > -3(1-w)(1+w).
\end{equation}
Substituting the CPL parametrization into this equation, and taking $a=1$ to correspond
to the present, we see that this translates into the limit
\begin{equation}
\label{wbound2}
w_a < 3(1-w_0^2),
\end{equation}
in order that the field not be rolling uphill today.
This limit is not a stringent bound, like Eq. (\ref{wbound}), so we will explore the parameter range outside
of it, but we note that such models seem somewhat less plausible for the reasons outlined above.

We now determine $V(\phi)$ for a range of values for $w_0$ and $w_a$.  We normalize $V(\phi)$ to its
present day value, $V_0$, and we use the freedom to shift $\phi$ by an arbitrary constant to
take $\phi=0$ to correspond to the present.  Because the CPL parametrization represents a heuristic fit to
the data, we will not require it to hold at arbitrarily early epochs, but only between some initial scale
factor $a_i$ and the present.  The choice for $a_i$ is somewhat arbitrary, but we will take $a_i = 0.3$
in what follows, so that all of the current supernova data lie in the interval $a_i \le a \le 1$.  The results
for $V(\phi)$ are shown in Figs. 1-2.  Note that for the particular case of $w_a=0$ (so that $w = w_0$), there
is an exact solution to Eqs. (\ref{V(a)}) and (\ref{phi(a)}), namely $V(\phi)$ is a power of a hyperbolic sine
\cite{UM,Rubano}.

\begin{figure}[htb]
\centerline{\epsfxsize=6truein\epsffile{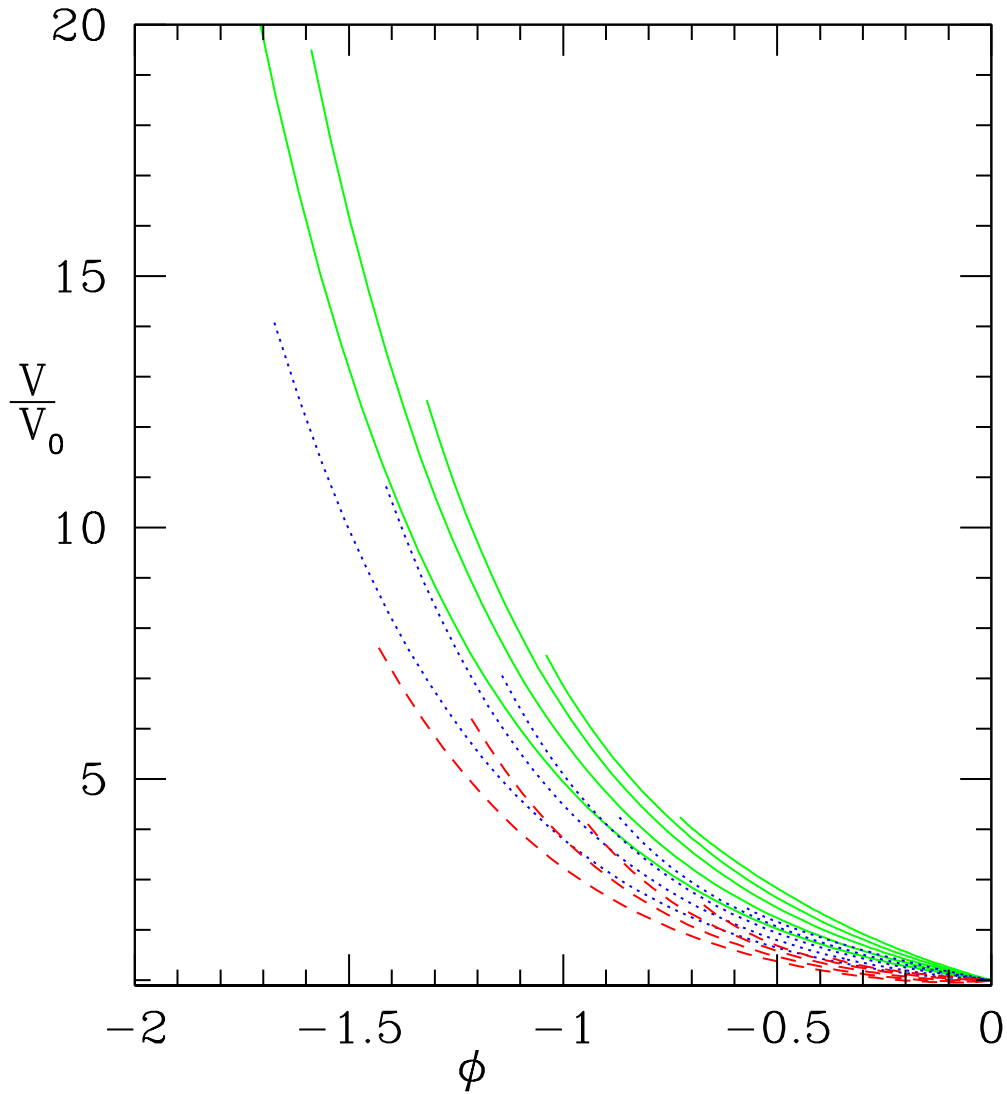}}
\caption{For quintessence models described by the CPL parametrization, the scalar
field potential $V(\phi)$ normalized
to its present day value, $V_0$, is plotted as a function of
$\phi$, for $w_0 = -0.9$ (dashed, red), and
$w_a =$ (top to bottom) 0, 0.5, 1, 1.5, 1.9; $w_0 = -0.6$ (dotted, blue) and 
$w_a =$ (top to bottom) $-0.5$, 0, 0.5, 1.0, 1.5; $w_0 = -0.3$ (solid, green) and
$w_a =$ (top to bottom) $-1.0$, $-0.5$, 0, 0.5, and 1.0, where we restrict the scale
factor $a$ to lie in the range $0.3 \le a \le 1$, and $\phi$ is translated such that
its present-day value is $\phi_0 = 0$.}
\end{figure}

\begin{figure}[htb]
\centerline{\epsfxsize=6truein\epsffile{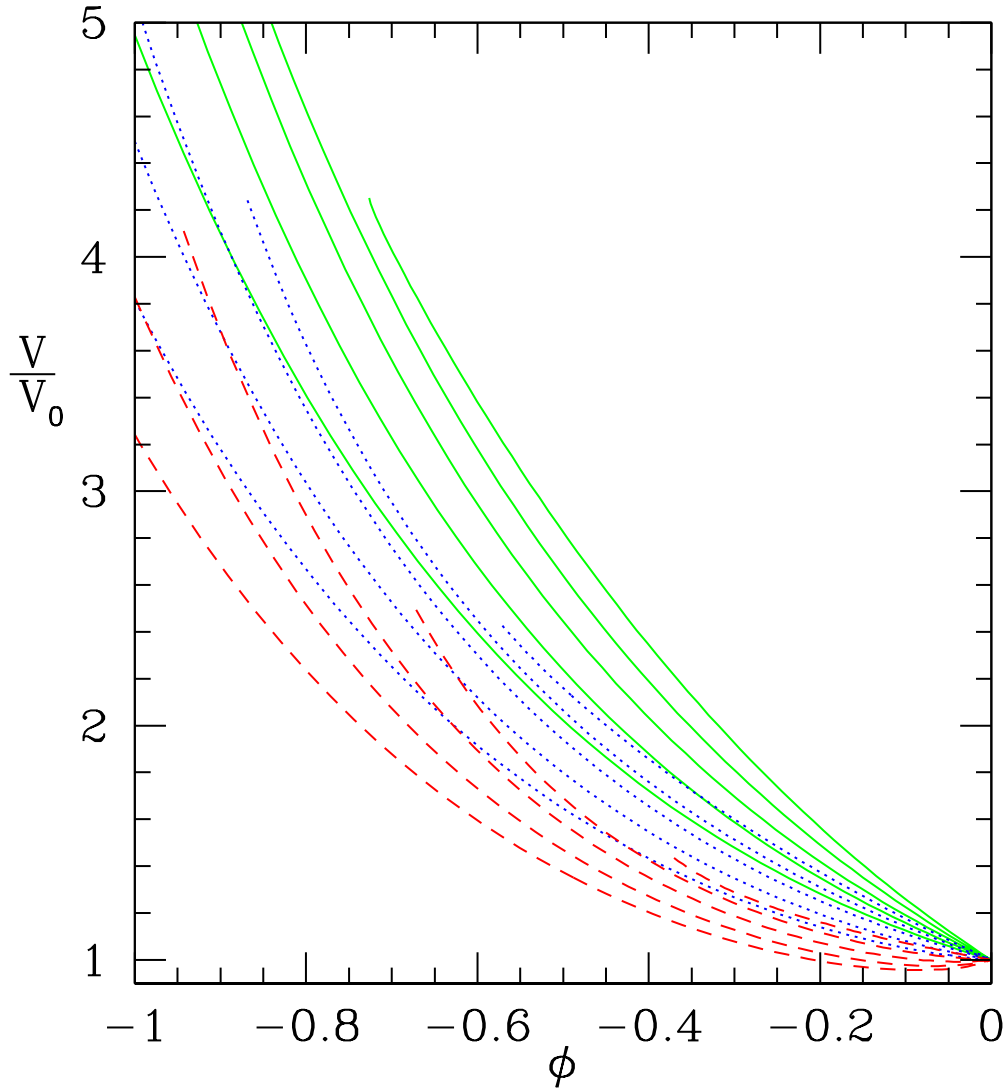}}
\caption{Expanded view of the region of Fig. 1 near $\phi=0$.}
\end{figure}

It is clear from Figs. 1-2 that the potentials corresponding to the CPL parametrization are all quite
similar to each other. In particular, fixing $w_0$
and varying $w_a$ produces a family of functions that are nearly indistinguishable, and the freezing
($w_a > 0$) and thawing ($w_a < 0$) models correspond to potentials which appear very similar.
(Note
that some of the curves with $w_0 = -0.9$ violate the bound in Eq. (\ref{wbound2}) and correspond
to scalar fields rolling uphill at the present; this can be seen in Fig. 2.)

All of the potentials in Figs. 1-2 appear to be convex functions ($d^2 V/d\phi^2 > 0$).  We can test
this hypothesis
using the direct expression for $d^2 V/d\phi^2$ from
Ref. \cite{Linderpaths} (see also Ref. \cite{Caldwell}), namely
\begin{equation}
\label{Vpp}
\frac{1}{H^2}\frac{d^2 V}{d\phi^2} = (2+3w+q/2)\frac{w^\prime}{1+w} + \frac{1}{4} \left(\frac{w^\prime}{1+w}\right)^2
- \frac{w^{\prime\prime}}{2(1+w)} + \frac{3}{4}(1-w)(5+3w+2q).
\end{equation}
Eq. (\ref{Vpp}) corrects a typo in the first term of Eq. (46) in Ref. \cite{Linderpaths} (E.V. Linder,
private communication).
Here the prime denotes the derivative with respect to $\ln a$, and $q$ is the deceleration parameter,
which, for a universe containing matter and a quintessence component with equation of state parameter
given by Eq. (\ref{CPL}), is
\begin{equation}
\label{q}
q = \frac{1}{2} + \frac{3}{2}\left(\frac{w_0 + (1-a) w_a}{1+(\rho_{M0}/\rho_{\phi0})
a^{3(w_0+w_a)}e^{3w_a(1-a)}}\right).
\end{equation}
Taking $0.3 \le a \le 1$, we scan over $w_0$ and $w_a$ and use Eqs. (\ref{Vpp})-(\ref{q}) to determine
the sign of $d^2 V/d\phi^2$.  We find that $d^2 V/d\phi^2 > 0$ except for two regimes. For
small values of $a$, $d^2 V/d\phi^2 < 0$ for $w_a > 1.5$, which is outside of the parameter
range explored in this paper.  The other case for which $d^2 V/d\phi^2 < 0$ is for $w_0
\le -0.9$,
$w_a$ small and negative, and $a$ near 1.  We will examine this regime in more detail shortly. 

Figs. 1-2 show that the CPL parametrization maps onto a very narrow range of functions $V(\phi)$.  However, this result
seems paradoxical,
since previous work \cite{ScherrerSen,Linderlinear1,Linderlinear2} suggests that a wide range of functional forms
for $V(\phi)$ are well-approximated by the CPL parametrization.  The key to resolving this apparent
contradiction is to note that these previous studies did not indicate that quintessence models map
to the full $w_0$, $w_a$ plane.  Instead, Refs. \cite{ScherrerSen,Linderlinear1,Linderlinear2} show that thawing quintessence maps
to a one-dimensional subspace of this plane, namely, the line defined by the relation
\begin{equation}
\label{linear}
w_a = \kappa (1+w_0),
\end{equation}
with $\kappa$ variously estimated to be in the range from $-1.6$ to $-1.5$.

To understand this result, we assume the validity of Eq. (\ref{linear}) and derive the corresponding
forms for $V(\phi)$ that
satisfy it.  These forms for $V(\phi)$ are shown in Fig. 3 for $\kappa$ ranging from $-0.5$ to $-3$.  Note that
in this case, we choose $a_i$ to be the value of $a$ for which $w(a) = -1$; this is the earliest allowed
initial value for $a$ for which this parametrization can be valid.  This yields $a_i = 1+1/\kappa$, independent
of $w_0$ and $w_a$.

\begin{figure}[htb]
\centerline{\epsfxsize=6truein\epsffile{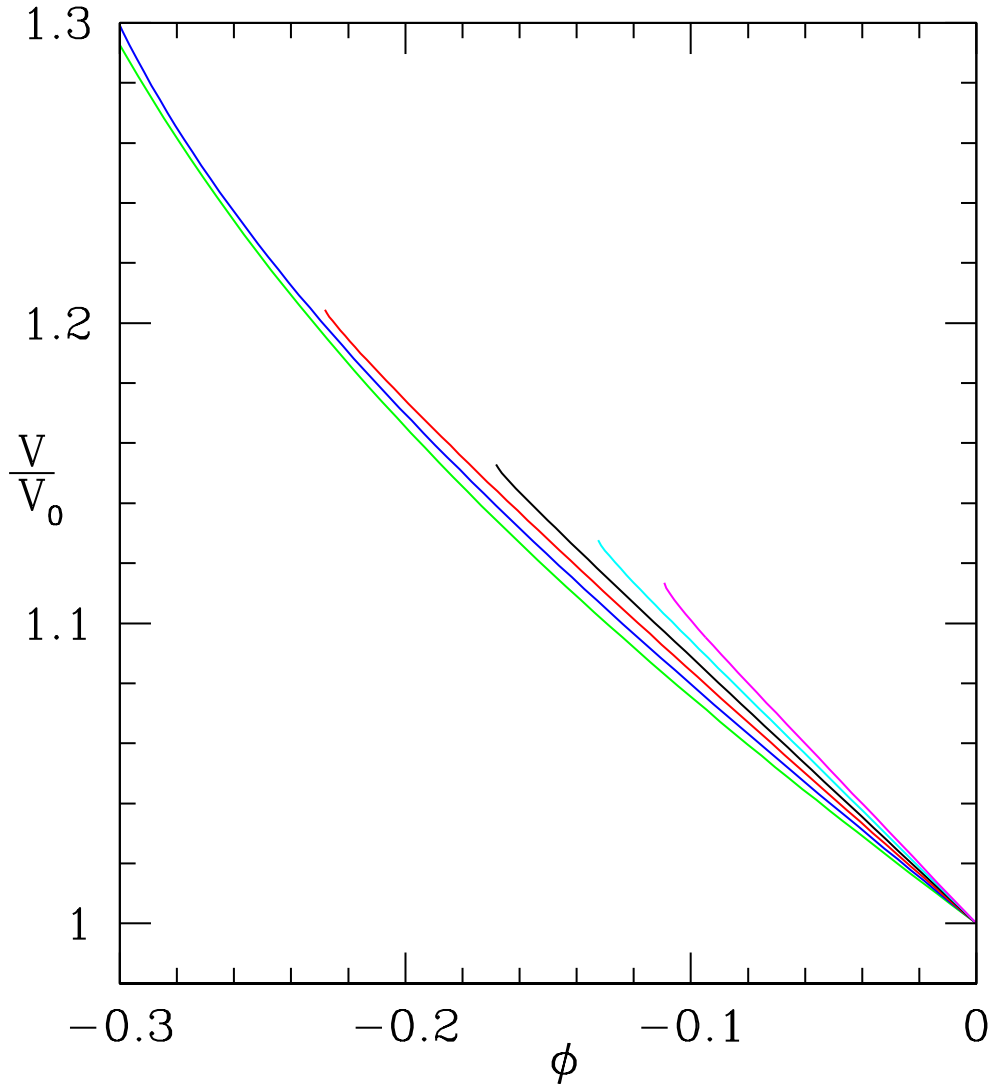}}
\caption{As Fig. 1, for $w_a = \kappa (1+w_0)$, with $w_0 = -0.9$ and $\kappa =$(bottom to top)
$-0.5$ (green), $-1$ (blue), $-1.5$ (red), $-2.0$ (black), $-2.5$ (cyan), $-3.0$ (magenta),
where $\phi$ is translated such that its present-day value is $\phi_0 = 0$, and $a_i$ is
the smallest value of $a$ for which $w(a) \ge -1$.}
\end{figure}

These forms for $V(\phi)$ become nearly linear for $\kappa = -1.5$,
while diverging from linear behavior on either side of this value.  This suggests that the particular choice
of Eq. (\ref{linear}) corresponds to a nearly linear potential.  Further, when $w$ is close to $-1$, (which
is always the case when $w_0 \approx -1$ and $w_a < 0$),
$\phi$ rolls only a short distance down the potential, and any sufficiently smooth potential will appear
to be linear on small enough scales.  This explains why thawing potentials with $w$ near $-1$ are described so well by
Eq. (\ref{linear}).

Further, Fig. 3 corresponds precisely to the potentials for which $V(\phi)$ is concave ($d^2 V/d \phi^2 < 0$) for $a$ near $1$.  This
corresponds to $\phi$ near $0$ in Fig. 3.  It is clear from this figure that $d^2 V/d \phi^2$ is negligibly small
in these cases.  Thus, we can conclude that for the parameter ranges examined in this paper, $d^2 V/d\phi^2$ is never large
and negative.  Thus, it makes sense that $w(a)$ for the hilltop models examined in Ref. \cite{ds1} 
(for which $d^2 V/d\phi^2$ is large and negative) strongly diverges from linear evolution in $a$.  This does {\it not}
mean that the CPL parametrization is a poor fit for all models with $d^2 V/d\phi^2 < 0$.  If $d^2 V /d\phi^2$
is sufficiently small, then our argument from the previous paragraph can apply,
yielding evolution well-described by the CPL parametrization and Eq. (\ref{linear}) \cite{ScherrerSen,Linderlinear2,ds1}.

Thus, we conclude that over most of the range in $w_0$ and $w_a$, the CPL parametrization maps onto a very narrow
form of behavior for $V(\phi)$, while a narrow range in the $w_0$, $w_a$ parameter space (given by Eq. \ref{linear}) maps to a wide
range of quintessence models, namely thawing quintessence with $w$ near $-1$.  However, it is precisely these
models that appear to be allowed by current observations, as emphasized by Ref. \cite{Linderlinear2}.

\section{Barotropic models}

In this section, we examine barotropic models, for which the pressure is a fixed function of the density:
\begin{equation}
\label{baro}
p = f(\rho).
\end{equation}
Particular models of this form include
the Chaplygin gas \cite{Kamenshchik,Bilic} and 
the generalized Chaplygin gas \cite{Bento}, the linear equation of state 
\cite{linear1,linear2} and the affine equation of state \cite{AB,Quercellini}  
(note these are actually the same model), the quadratic 
equation of state \cite{AB}, and the Van der Waals equation of state
\cite{VDW1,VDW2}.  A general study of the properties of barotropic models for dark energy
was undertaken in Ref. \cite{LinderScherrer} and further extended in Ref. \cite{Bielefeld}.  Note
that there is a simple mapping between the barotropic models discussed here and kinetic $k$-essence
models \cite{LinderScherrer}, so the results presented here can be extended in a straightforward way
to the latter set of models.

Ref. \cite{LinderScherrer} provided two reasonable constraints on the behavior of $w(a)$ in
barotropic models.  To prevent instabilities, the sound
speed $c_s^2 = dp/d\rho$ must obey $c_s^2 \ge 0$, while causality
requires $c_s^2 \le 1$.  While each of these limits translates into a constraint on $w(a)$,
it is easier in this case to derive the form for $p$ as a function of $\rho$ and
then apply the direct limits on $dp /d\rho$:
\begin{equation}
\label{barolimits}
0 \le \frac{dp}{d\rho} \le 1.
\end{equation}

If the equation of state for the barotropic fluid  is given by Eq. (\ref{CPL}), then the expression
for the density $\rho$ given by Eq. (\ref{rhoCPL}) is valid as well, i.e.,
\begin{equation}
\label{rhoCPLagain}
\rho = \rho_0 a^{-3(1+w_0+w_a)} e^{3w_a(a-1)},
\end{equation}
Further, the pressure will be
given by $w \rho$, so
\begin{equation}
\label{pCPL}
p = \rho_0 [w_0 + (1-a)w_a] a^{-3(1+w_0+w_a)} e^{3w_a(a-1)}
\end{equation}
Eqs. (\ref{rhoCPLagain}) and (\ref{pCPL}) together provide an implicit expression for $f(\rho)$
in Eq. (\ref{baro}).  As in the previous section, we will not require the CPL parametrization to hold
all of the back to $a=0$, but only for $a_i \le a \le 1$, where we take $a_i=0.3$.  In Figs. 4-5, we plot $p/\rho_0$
as a function
of $\rho/\rho_0$ for the scale factor lying in this interval.

\begin{figure}[htb]
\centerline{\epsfxsize=6truein\epsffile{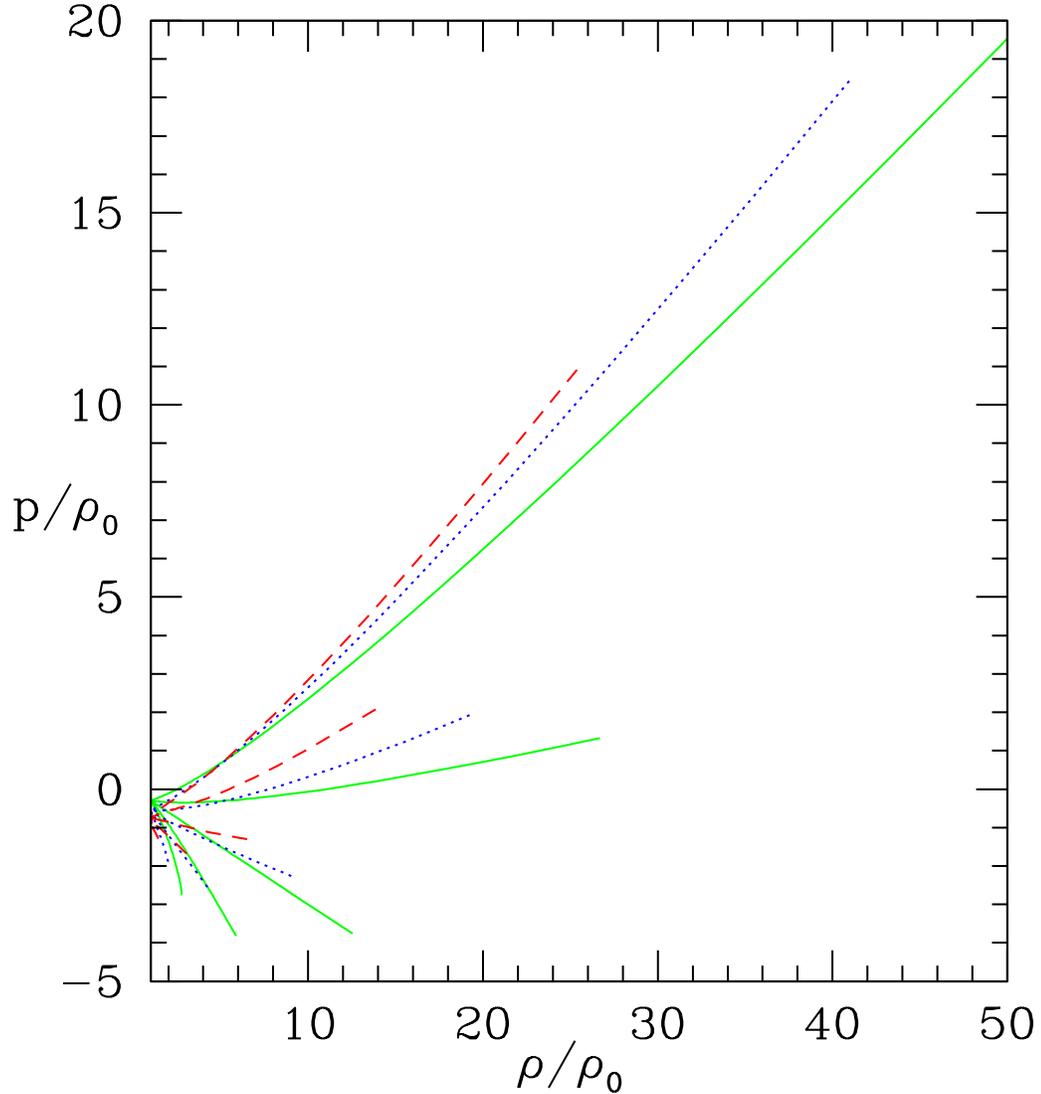}}
\caption{For barotropic models described by the CPL parametrization,
the pressure $p$, given as $p/\rho_0$, is plotted as a function of the density $\rho$, given as
$\rho/\rho_0$, where $\rho_0$ is the present-day dark energy density,
for $w_0 = -0.9$ (dashed, red), and
$w_a =$ (bottom to top) 0, 0.5, 1, 1.5, 1.9; $w_0 = -0.6$ (dotted, blue) and 
$w_a =$ (bottom to top) $-0.5$, 0, 0.5, 1.0, 1.5; $w_0 = -0.3$ (solid, green) and
$w_a =$ (bottom to top) $-1.0$, $-0.5$, 0, 0.5, and 1.0, where we restrict the scale
factor $a$ to lie in the range $0.3 \le a \le 1$.}
\end{figure}

\begin{figure}[htb]
\centerline{\epsfxsize=6truein\epsffile{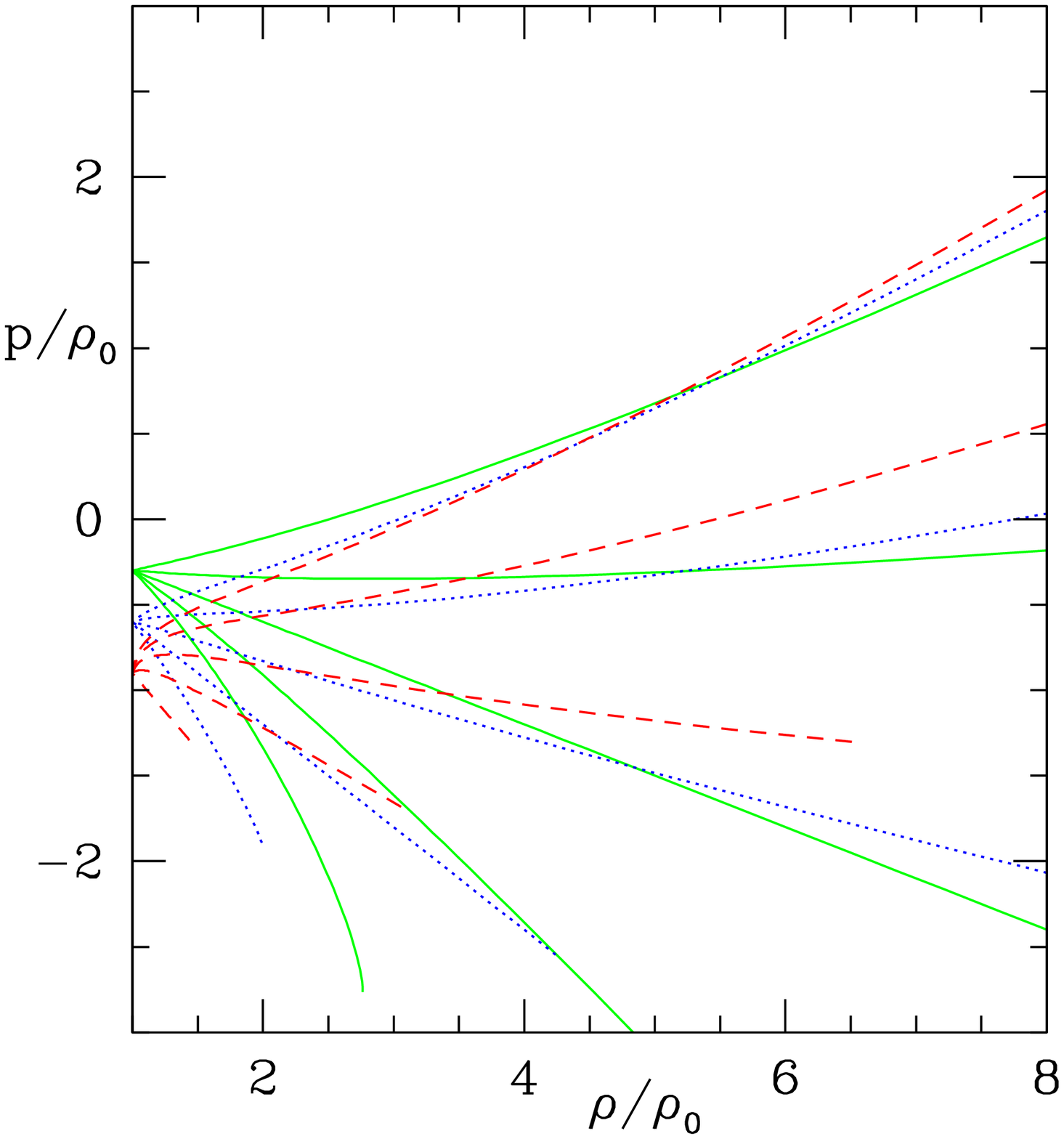}}
\caption{Expanded view of the region of Fig. 3 near $\rho/\rho_0=1$.}
\end{figure}

The behavior of $p$ as a function of $\rho$ for these barotropic models obeying the CPL
parametrization displays a much wider set of functional behaviors than is the case for
$V(\phi)$ in the quintessence models.  Note, however, that
all of the curves converge on the same value of $p/\rho_0$ ($=w_0$) at $\rho/\rho_0 = 1$, corresponding to $a=1$.

We can apply the limits from Ref. \cite{LinderScherrer} given by Eq. (\ref{barolimits})
directly to the models displayed in Fig. 4.  It is clear that causality ($dp/d\rho \le 1$)
is satisfied for all of the models under consideration.  However, the requirement
that $dp/d\rho \ge 0$ is violated by some of these models.  From Figs. $4-5$,
it is clear that causality requires relatively large positive values of $w_a$ (i.e.,
freezing models rather than thawing models).

We can make a stronger statement about the allowed forms for $f(\rho)$ and their dependence on
$w_0$ and $w_a$. Eq. (\ref{CPL}) can be used to express the scale factor as a function of $w$, $w_0$, and
$w_a$:
\begin{equation}
a = \frac{w_a + w_0 - w}{w_a},
\end{equation}
and this expression can be substituted
into equation (\ref{rhoCPLagain}) to give
\begin{equation}
\label{implicitrho}
\rho = \rho_0 |w_a|^{3(1+w_i)} e^{-3w_a}|w_i - w|^{-3(1+w_i)} e^{3(w_i-w)},
\end{equation}
where we define $w_i$ to be the value of $w$ at $a=0$ in the CPL approximation,
namely, $w_i = w_0 + w_a$.
Note that Eq. (\ref{implicitrho}) is an implicit equation for $p(\rho)$, since $w = p/\rho$.
Now define
\begin{eqnarray}
\widetilde \rho = \rho/\rho_C,\\
\widetilde p = p/\rho_C,
\end{eqnarray}
with the constant density $\rho_C$ given by
\begin{equation}
\rho_C = \rho_0 |w_a|^{3(1+w_i)} e^{-3w_a}
\end{equation}
Then Eq. (\ref{implicitrho}) becomes
\begin{equation}
\widetilde \rho = |w_i - \widetilde p/\widetilde \rho|^{-3(1+w_i)} e^{3(w_i-\widetilde p/\widetilde \rho)}.
\end{equation}
From this equation, it is clear that the expression for $\widetilde p$ as a function
of $\widetilde \rho$ depends only on $w_i$ and not on $w_a$ and $w_0$ independently.  Writing
this function as $\widetilde p = g(\widetilde \rho)$, we then have
\begin{equation}
p/p_C = g(\rho/\rho_C)
\end{equation}
where the function $g$ depends only on $w_i$.
Thus, aside from an overall multiplicative constant in both $\rho$ and $p$ (which is the same
for $\rho$ and $p$, so that $w$ is unaltered), $f(\rho)$ in Eq. (\ref{baro})
depends
only on $w_i$, and not on $w_a$ and $w_0$ separately.  

\section{Conclusions}
Lacking a priori knowledge of the actual physical model underlying the accelerated expansion of the universe,
it is impossible to make blanket statements about the utility of
of the CPL parametrization for approximating such models.  However, we can make statements about individual dark energy models;
the implicit question is the
extent to which a linear parametrization of $w$ provides useful information about the parameters of a given
physical model.

For the case of quintessence, we have seen that the full parameter space of the CPL parametrization maps
to only a fairly narrow form of functional behavior for $V(\phi)$.
However, a one-dimensional subset of CPL
parameter space, namely $w_a = \kappa (1+w_0)$, with $\kappa \approx -1.6$ - $-1.5$ corresponds to a wide
range of functional forms for $V(\phi)$.  This issue has been noted by Linder \cite{Linderlinear2}, who pointed
out that such models are better fit by the one-parameter model of Eq. (\ref{linear}) than by the full CPL parametrization.
Further, Ref. \cite{Linderlinear2} introduced a more accurate two-parameter approximation, using Eq. (\ref{linear})
as a starting point, that better fits the behavior of thawing quintessence models.  Our results provide further
evidence in favor of this approach.

We note a similar situation for barotropic models.  For these models,
the CPL parametrization does a much better job of capturing the full range of possible
functional forms for $p = f(\rho)$  (the pressure as a function of density).
However, we have shown explicitly that the barotropic models that correspond exactly to the CPL
parametrization belong to a one-dimensional subset
of the full CPL parameter space; namely, they are fully determined (up to a multiplicative constant) by the value
of $w_i = w_0 + w_a$.  Thus, the full CPL approximation may not be the most useful way to characterize such models.

Considered as a straightforward linear approximation for the equation of state parameter as a function of the expansion factor,
the CPL parametrization is undeniably useful, particularly when attempting to detect deviations from the $\Lambda$CDM
model.  However, as a probe of physical models which might correspond to non-$\Lambda$CDM behavior (or at least the types
of physical models considered here), the CPL parametrization is not optimal, and, at least for the
case of quintessence, it seems more appropriate to move toward the kind of parametrization discussed
in Ref. \cite{Linderlinear2}.
It would, of course, be useful to extend this study to some of the other classes of models for dark energy that have been
proposed in the literature.

\section{Acknowledgments}
The author thanks E.V. Linder for a helpful critique of an early draft of this paper,
and particularly for pointing out Eq. (\ref{Vpp}).
R.J.S. was supported in part by the Department of Energy
(DE-SC0011981).

\end{document}